\documentclass[a4paper,twoside,twocolumn,english,amssymb,aps,prb,reprint,twocolumn,superscriptaddress,floats,showkeys,showpacs]{revtex4-1}
\usepackage{graphicx,amsmath}
\usepackage{epstopdf}
\usepackage{color}
\usepackage[T1]{fontenc} 
\usepackage[utf8]{inputenc}

\begin{document}
\title{From Quantum Dots to Quantum Dashes: Excitonic Spectra of Highly Elongated InAs/InP Nanostructures}
\author{M. Zieli\'nski}
\email{mzielin@fizyka.umk.pl}
\affiliation{Institute of Physics, Faculty of Physics, Astronomy and Informatics, Nicolaus Copernicus University, Grudziadzka 5, 87-100 Torun, Poland}

\begin{abstract}
A transition from a cylindrical quantum dot to a highly elongated quantum dash
is theoretically studied here with an atomistic approach combining
empirical tight binding for single particle states and
configuration interaction method for excitonic properties.
Large nanostructure shape anisotropy leads to a peculiar trend of the bright exciton splitting, which at certain point
is quenched with further shape elongation, contradicting predictions of simplified models.
Moreover strong shape elongation promotes pronounced optical activity of the dark exciton,
that can reach substantial $1\%$ fraction of the bright exciton intensity without application of any external fields.
An atomistic calculation is augmented with a elementary phenomenological model 
expressed in terms
of light-hole exciton add-mixture increasing with the shape deformation.
Finally, exctionic complexes $X^-$, $X^+$, and $XX$ are studied as well and the correlations due to presence of higher excited states are identified as a key-factor affecting excitonic binding energies and the fine structure.

\end{abstract}
\maketitle

\section{Introduction}
Self-assembled InAs/InP nanostructures are promising quantum emitters at $1.3$ or $1.55$~$\mu$m telecom wavelengths.~\cite{Anufriev,yuan2011controlling,benyoucef2013telecom,kors2017telecom}
Potential usage of these systems involves single photons and entangled photons generation,~\cite{Michler2282,PhysRevLett.86.1502,PhysRevLett.84.2513,Stevenson2006} with applications in quantum information and quantum communication.~\cite{Knill2001,Waks2002,RevModPhys.79.135}
The realm of InAs/InP nanostructures is rich and varies from more conventional cylindrical self-assembled,~\cite{williams2001controlling,kors2018telecom} and nanowire quantum dots~\cite{dalacu-selective,Versteegh2014,maaike-strain} to rather unconventional semiconductor nanostructures with characteristic large in-plane elongation, known as quantum dashes.~\cite{dery2004nature,sauerwald2005size,ReithmaierQDash1,ReithmaierQDash2,khan2014self,musial2012carrier,musial2014toward,gawelczyk2017exciton,dusanowski2017confinement,musial2014phonon,jung2018effect}
Quantum dashes have demonstrated their potential for utilization in e.g. lasers and amplifiers~\cite{ReithmaierQDash1,ReithmaierQDash2,azouigui2009optical} or single photon emitters.~\cite{dusanowski2013exciton,dusanowski2014single}
Quantum dashes show also significant tuning capabilities e.g. by embedding in photonic mesas~\cite{mrowinski2018photonic} as well as demonstrate several unique properties such a suppression of phonon-induced decoherence.~\cite{musial2014phonon}

Recently, by application of external magnetic field, quantum dashes~\cite{mrowinski2015magnetic} have demonstrated their capability to systematically reduce the bright exciton fine structure splitting,~\cite{bayer-eh} below the natural line width of emission, which makes the quantum dashes prospective as a source of polarization entangled photon pairs from biexciton-exciton cascade.~\cite{PhysRevLett.84.2513} The reduction of the fine structure splitting, aiming for entanglement generation, in an important scientific topic, and significant efforts have been made to achieve this goal including, among many others, post-growth annealing,~\cite{langbein2004control,young2005inversion} spectral filtering, ~\cite{akopian2006entangled} sample selection,~\cite{hafenbrak2007triggered,olbrich2017polarization} 
growth of high symmetry structures,~\cite{singh-bester-eh,karlsson,Dupertuis.PRL.2011,Versteegh2014,kors2018telecom} the application of external electric,~\cite{kowalik2005influence,bennett2010electric} magnetic,~\cite{Stevenson2006,stevenson2006magnetic,gerardot2007manipulating} and strain fields.~\cite{seidl2006effect,ding2010tuning,trotta2012universal} 
It is thus of practical importance to study, as done in this work, the bright exciton fine structure splitting in quantum dot-quantum dash systems and see how it evolves with deformation for high-aspect ratio structures. At the other end of excitonic spectra are the dark exciton states, which recently gained an attention as potential long-lived, yet optically addressable quantum bits~\cite{PhysRevB.92.201201,PhysRevX.5.011009,Zielinski.PRB.2015} or could be utilized as auxiliary (metastable) states for the time-bin entanglement generation scheme.~\cite{PhysRevLett.94.030502,weihs2017time}
For the same reasons the biexciton spectra is particularly interesting as well, and hence both the dark exciton and several excitonic complexes, such as the biexciton, are studied in this paper in detail.

Apart from any practical utilization quantum dashes are intriguing from a basic science point of view.
For example, it is curious how single particle and many body properties evolve from that characteristic for a cylindrical quantum dot, by gradual deformation of nanostructure's shape to a high-aspect ratio, deformed quantum dash.
Such observation is typically not directly possible in the experiment, yet it is fully attainable in a theoretical study, which by its nature focuses on individual, single quantum systems of well defined size, shape and composition.
Further, often the experiment is pestered by uncertainties due to inhomogeneous broadening in the ensemble studies~\cite{dash-mrowinski} or unavoidable alloy randomness effects originating from the specifics of the epitaxial growth,~\cite{dash-mrowinski,Zielinski-natural} in particular growth 
on mixed composition substrates (e.g. InGaAlAs~\cite{marynski2013electronic,jung2018effect}). 
A detailed knowledge (chemical composition, intermixing, actual dimensions) of nanostructure's morphology, is often very much complicated, if not impossible to obtain.~\cite{dery2004nature,sauerwald2005size,khan2014self,Mlinar-inverse}
Theoretical simulation can to a certain degree filter out these difficulties, 
and focus not on the direct comparison with a particular quantum dash or quantum dot experiment, but rather an analysis of general trends with shape elongation. 
Utilization of accurate (e.g. atomistic) approach should however give hope to produce results consistent and supporting experimental findings.
Moreover theoretical studies should also give an insight used next for e.g. intentional tailoring of high-aspect ratio nanostructures to match spectral features demanded in the broad field of nanophotonics.~\cite{salter2010entangled,santori2002indistinguishable,gerard1998enhanced,peter2005exciton}

The focus of this paper is on the details of the InAs/InP quantum dot-quantum dash single exciton spectra and in particular the fine structure splitting, as well as the bright and the dark exciton optical activity, all calculated as a function of shape deformation from high rotational shape symmetry to high-aspect ratio elongation. Spectra of several excitonic complexes, the negatively and positively charged exciton and the biexciton, are studied here as well. For all these cases, this article underlines the key role of correlation effects due to presence of higher levels (configuration mixing) affecting significantly all major features of quantum dashes excitonic spectra.

\subsection{System and methods}
\begin{figure}
  \begin{center}
  \includegraphics[width=0.45\textwidth]{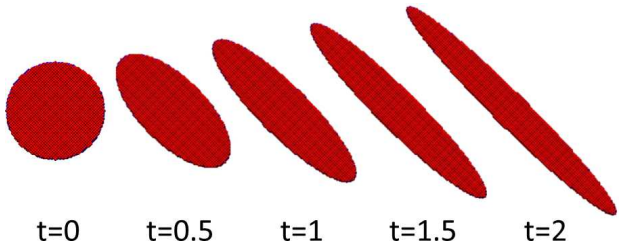}
  \end{center}
  \caption{Schematics (top-view) of InAs nanostructure shape deformation along [1\underline{1}0] crystal axis as a function \textit{t}, for several arbitrarily chosen values. InAs wetting layer below the nanostructure and the surrounding InP barrier is omitted for clarity.}
  \label{scheme}
\end{figure}

In this work I study InAs nanostructures elongated from an ideal disk-shape quantum dot to a highly-elongated (nanorod-like) elliptical quantum dash.
The height of all deformed nanostructures is kept fixed and equal to $3$~nm. The diameter in a fully cylindrical case is $20.6$~nm (radius $r=10.3$~nm). These dimensions are typical for self-assembled quantum dots and quantum dashes.~\cite{khan2014self}
In our previous work~\cite{dash-mrowinski} we have studied effects of nanostructure size and composition on quantum dashes spectra, here
focus on the role of lateral deformation for pure InAs/InP system.
The anisotropy is applied as by elongating the system along $[1\underline{1}0]$ axis and shrinking it in the perpendicular $[110]$ axis at the same time, such that the base field (and the overall dot volume) is kept constant.
The elongation axis was chosen as $[1\underline{1}0]$, consistent with experimental findings.~\cite{mrowinski-pol}
The elongation is governed by anisotropy~\cite{kadantsev-eh,zielinski-elong} parameter $t$ with a longer (major) axis length changing as $X=r\times\left(1+t\right)$
and a shorter (minor) axis length given by $Y=r/\left(1+t\right)$. We consider $t$ to vary from $0$ to $2.0$ (Fig.~\ref{scheme}), and thus the aspect ratio ($X/Y$) is given as $(1+t)^2$ and reaches $9$ for the highest deformation considered. 
The nanostructure is located on a $1$ lattice constant ($2$ monolayers) thick ($0.6$~nm) InAs wetting layer~\cite{rudno2006photoreflectance} and embedded in InP barrier.
Since $t$ was varied from $0$ to $2.0$ with a $0.1$ step the calculations we performed for total of $21$ different systems.

The calculation starts with finding atomic positions that minimize total elastic energy.
This is done by using the valence force field method of Keating~\cite{keating,martin}
with a cylindrical computational box containing total of over $32$~million atoms,
and the minimization of strain energy performed using the conjugate gradient method.~\cite{jaskolski-zielinski-prb06}
The valence force method is described in more detail in Refs.\cite{pryor-zunger,saito-arakawa} and in my  previous
papers.~\cite{jaskolski-zielinski-prb06,zielinski-prb09,zielinski-including,zielinski-vbo}
Next, from atomic positions the piezoelectric potential is calculated by accounting for both linear and quadratic~\cite{PhysRevLett.96.187602,PhysRevB.74.081305,PhysRevB.84.195207} terms.

Then the single particle spectra of electrons and holes are obtained with the empirical tight-binding method accounting for $d$-orbitals and spin-orbit interaction.~\cite{zielinski-including,zielinski-vbo}
The single-particle tight-binding Hamiltonian for the system of $N$ atoms and $m$ orbitals per atom can be written, in the language of the
second quantization, in the following form:~\cite{zielinski-prb09}
{\begin{multline}
  \hat{H}_{TB} =
  \sum_{i=1}^N \sum_{\alpha=1}^{m}
       E_{i\alpha}c_{i\alpha}^+c_{i\alpha}
+  \sum_{i=1}^N \sum_{\alpha=1,\beta=1}^{m}
       \lambda_{i\alpha,\beta}c_{i\alpha}^+c_{i\beta}\\
+   \sum_{i=1}^N \sum_{j=1}^{near. neigh.} \sum_{\alpha,\beta=1}^{m}
       t_{i\alpha,j\beta}c_{i\alpha}^+c_{j\beta}
\label{tb-ham}
\end{multline}}
where $c_{i\alpha}^+$ ($c_{i\alpha}$) is the creation (annihilation) operator of a carrier on the (spin-)orbital $\alpha$ localized on the site
$i$, $E_{i\alpha}$ is the corresponding on-site (diagonal) energy, and $t_{i\alpha,j\beta}$ describes the hopping (off-site, off-diagonal)
of the particle between the orbitals on (four) nearest neighboring sites.
The summation $i$ goes over all atoms, whereas the summation over $j$ goes over the four nearest neighbors only.
$\alpha$ is a composite (spin and orbital) index of the on-site orbital, whereas $\beta$ is a composite index of the neighboring atom orbital.
Coupling to further neighbors is thus neglected, whereas $\lambda_{i\alpha,\beta}$ (on-site, off-diagonal) accounts for the
spin-orbit interaction following the description given by Chadi\cite{chadi-so-in-tb} and including the contributions from atomic $p$ orbitals.

Here I use the $sp^3d^5s^*$ parametrization of Jancu~\cite{jancu}.
The tight-binding calculation is effectively performed on a smaller domain than the valence force field calculation,~\cite{lee-boundary,zielinski-multiscale}
yet still the number of atoms in the tight-binding computational box exceeds $1.3$~million presenting a substantial numerical problem.
More details of the $sp^3d^5s^\star$ tight-binding calculation were discussed thoroughly in our earlier papers.~\cite{zielinski-including,zielinski-vbo,jaskolski-zielinski-prb06,zielinski-prb09}

Finally excitonic spectra are calculated with the configuration interaction method.
The Hamiltonian for the interacting electrons and holes can be written in second
quantization as:~\cite{michler}
{\begin{multline}
  \hat{H}_{ex} =  \sum_{i}E_i^ec_i^\dagger c_i+\sum_{i}E_i^hh_i^\dagger h_i  \\
  +\frac{1}{2}\sum_{ijkl}V_{ijkl}^{ee} c_i^\dagger c_j^\dagger c_k c_l
  +\frac{1}{2}\sum_{ijkl}V_{ijkl}^{hh} h_i^\dagger h_j^\dagger h_k h_l \\
  -\sum_{ijkl}V_{ijkl}^{eh,\text{dir}} c_i^\dagger h_j^\dagger h_k c_l
  +\sum_{ijkl}V_{ijkl}^{eh,\text{exchg}} c_i^\dagger h_j^\dagger c_k h_l
\end{multline}}
where $E_i^e$ and $E_i^h$ are the single particle electron and hole energies,
obtained at the single particle stage of calculations, and
$V_{ijkl}$ are Coulomb matrix elements (Coulomb direct and exchange integrals) calculated according to procedure given in Ref.~\cite{zielinski-prb09}.

Typically only $s$, $p$, and $d$ single particles shells are included in the configuration interaction calculation of quantum dot electronic spectra.~\cite{zielinski-prb09,zielinski-tallqd}
In this work I go beyond this approximation and account for the $f$-shell as well, resulting in total $20$ (with spin) electron and $20$ hole single particle states entering
many-body calculations and total of over $0.6$~million Coulomb direct and exchange integrals calculated over $1.3$~million atoms in the computational box.~\cite{zielinski-prb09,rozanski-zielinski}
Calculation of the multi-excitonic spectra produces thus a significant computational challenge and on a $192$-core computer cluster it takes about $72$-hours for all computational stages combined (i.e. strain, piezoelectricity, tight-binding and configuration interaction) for every single $t$ value, with the configuration interaction being by far the most time-demanding part.

\section{Single particle spectra and the excitonic ground state}

\begin{figure}
  \begin{center}
  \includegraphics[width=0.45\textwidth]{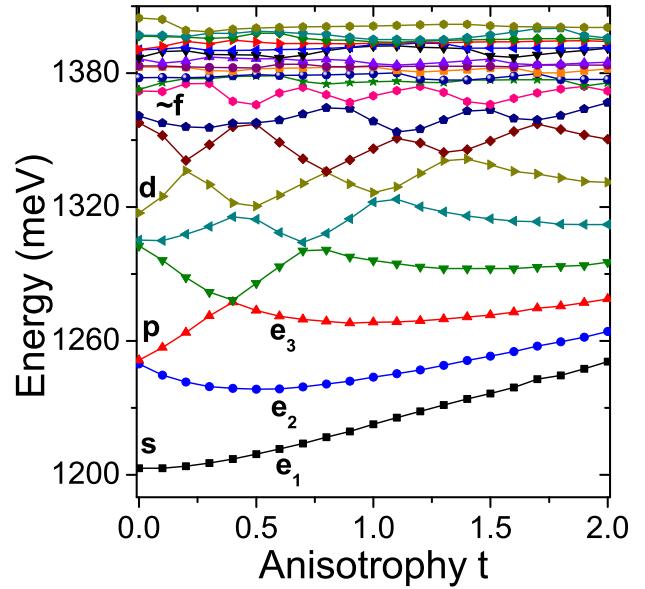}
  \end{center}
  \caption{Single particle electron levels as a function of nanostructure shape deformation \textit{t} along [1\underline{1}0] crystal axis.
   Lines connect levels ordered in energy. Three lowest states are marked as $e_1$, $e_2$, and $e_3$ correspondingly.}
  \label{electrons}
\end{figure}

Fig.~\ref{electrons} shows the evolution of several lowest electron levels as a function of increasing lateral shape anisotropy.
These levels evolve from the typical case of a cylindrical quantum dot with a characteristic $s$, $p$, $d$, ... shell structure~\cite{arek-book} to
a nanorod-like spectrum with nearly equidistant levels and no shells present.
In a simplified picture this could be understood as an evolution from a quasi-two-dimensional like confinement effectively described by a $2D$-harmonic oscillator~\cite{arek-book,michler} model to a quasi-one-dimensional system described by a $1D$-harmonic oscillator model, hence nearly equidistant spacings between levels.

Apart from above lowest electron levels experience a blue-shift in energy, with a pronounced increase of the ground electron state energy ($e_1$)
by $48$~meV for the largest considered deformation ($t=2$) as compared with the cylindrical case ($t=0$).
This happens despite keeping the nanostructure's volume fixed during the deformation, as mentioned above,
and is due to decreased confinement in the $[110]$ direction, which is perpendicular to the direction of the nanostructure deformation direction $[1\underline{1}0]$.
The $f$-shell seems to be weakly pronounced and at about $1380$~meV there is an apparent onset of closely spaced quasi-continuum states ($2D$-like wetting layer states).
Eight (16 with spin) lowest electrons states are typically well confined within the nanostructure with the percent of the wave-function contribution in the quantum dot region varying from about $78\%$ for the $e_1$ ground state to about $65\%$ for $e_8$, and for $t=0$. Higher energy states are weakly coupled with small $30-45\%$, yet non-negligible, contribution in the nanostructure volume.
Higher lying states have progressively increasingly larger content in the wetting layer and in the barrier region.
The elongation further reduces localization in the dot region which for $e_1$ drops from $78\%$ for $t=0$ to $69\%$ for $t=2$ and analogously for higher states.

\begin{figure}
  \begin{center}
  \includegraphics[width=0.45\textwidth]{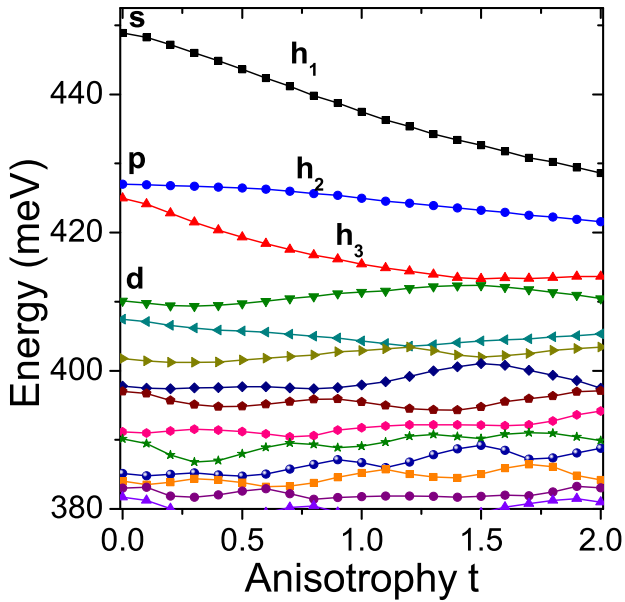}
  \end{center}
  \caption{Single particle holes levels as a function of nanostructure shape deformation \textit{t} along [1\underline{1}0] crystal axis.
   Lines connect levels ordered in energy. Please noted reversed ordering of hole levels with respect to electron levels
   The hole ground state is marked as $h_1$, the first excited hole states is $h_2$, etc.}
  \label{holes}
\end{figure}

Evolution of hole levels, as shown on Fig.~\ref{holes}, is somewhat more peculiar with a complicated spectra in the elongated case,
several apparent anti-crossings of excited levels, and decreasing\footnote{Please note the reversed order of hole levels with respect to the electron.}
energy of the ground hole $h_1$ level by about $20$~meV with elongation.
As typical for various nanostructures hole states have smaller energy levels spacings due to the larger effective mass of the hole. Hence for the non-elongated case ($t=0$), the hole $p-shell$ splitting appears larger than that of electron, whereas in fact both splittings are quite comparable,
with $1.63$~meV value for electrons and $2.1$~meV splitting for holes.
It should be noted that the hole $d$-shell is apparently less separated from the excited part of the spectra than in the electron case, the $f$-shell is
practically not visible, and that the confined hole shells smoothly transform into closely separated levels below about $385$~meV.
Yet, contrary to the electron, and again due to larger effective mass, even excited holes states are well confined in the nanostructure with above $95-96\%$ hole wave-function localization in the dot region for all $10$ ($20$ with spin) lowest ($h_1$-$h_{10}$) hole states considered in latter part of the text with respect to the excitonic calculations. Interestingly the shape-deformation has little (reduction on the order of at most $\%1$) effect on the ground hole state $h_1$ localization degree.

At the end of the single particle energies analysis, let me note that the ground and the first excited state ($s-p$) spacing decreases monotonically with deformation, dropping for electrons (Fig.~\ref{electrons}) from $47$~meV 
(for $t=0.0$) to $13$~meV (for $t=2.0$), and for holes (Fig.~\ref{holes}) from $22$~meV to $7$~meV correspondingly.

\begin{figure*}
  \begin{center}
  \includegraphics[width=0.95\textwidth]{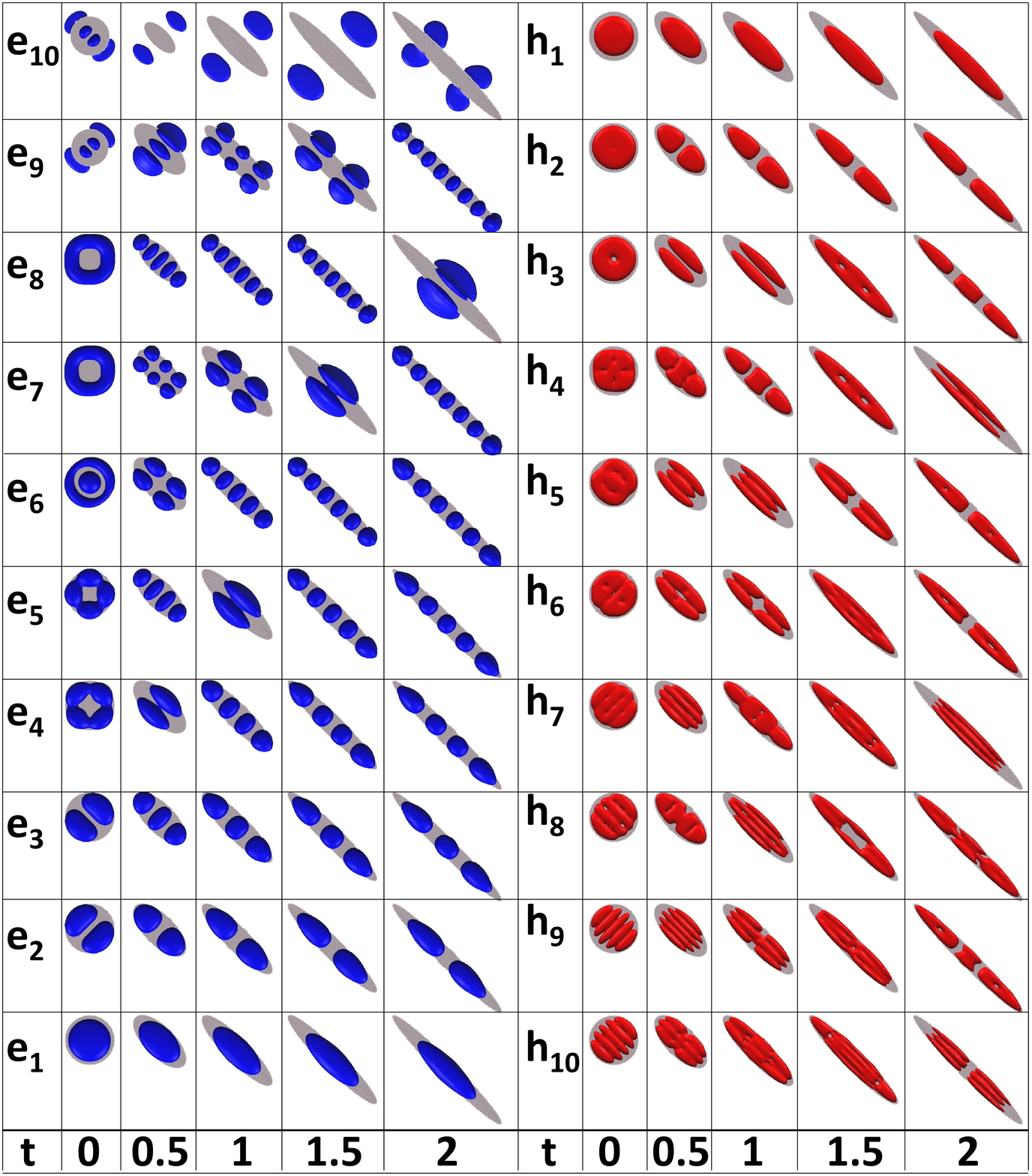}
  \end{center}
  \caption{Charge densities (top-view) corresponding to several ($10$) lowest single particle electron (left) and hole (right) levels as a function of nanostructure shape deformation \textit{t=0, 0.5, 1, 1.5, 2}.
  Please note reversed ordering of hole levels. 
  Some of these pictures were slightly re-scaled to fit the table.
  All iso-surfaces were plot using the same (arbitrary) charge-density constant value. 
  Grey areas mark the nanostructure region.
  Please thus note that some of these states are not bound within the nanostructure (e.g. $e_{10}$ for \textit{t=1}). Moreover several of these states leaks profoundly into the wetting layer (e.g. $e_{8}$ for \textit{t=2}).}
  \label{charge_dens}
\end{figure*}

Fig.~\ref{charge_dens} shows charge densities corresponding to single particles levels shown on Fig.~\ref{electrons} and Fig.~\ref{holes}. For electron states the transition from a cylindrical to a quasi-one-dimensional confinement is apparent and manifests itself in the charge density by an increasing number of nodal planes along the elongation axes. This is particularly well visible for highly-elongated cases (e.g. $t=2$) were the manifold of lowest electron states starts for a node-less ground states, then the first excited state has one nodal plane, the second excited states has two nodal planes, and so on.

This simple, intuitive picture is somewhat complicated by a presence of apparently delocalized states intruding in-between localized ones. This is well visible for example for $e_8$ and $t=2$. These particular state apparently origins from  $e_3$, one of the cylindrical ($t=0$) quantum dot $p-shell$ states, which evolves higher in energy with the shape-deformation due to a presence of a node in this state aligned with the longer axis of the nanostructure. Such near-delocalized states have significant fraction of their charge density ''leaking'' out of the nanostructure into the InAs wetting layer and finally into the InP barrier. 
Another example of an interesting behaviour is seen for higher excited states $e_9$ and $e_{10}$ for $t=0$ and $t=0.5$ cases. Here again these states have mixed character with a fraction of charge density in the nanostructure hybridized with the rest of it penetrating the wetting layer. Finally there are electron states shown here as well (e.g. $e_{10}$ for $t=1.5$), with no apparent density in the quantum dash area, but rather delocalized in the wetting layer.

The ground hole state charge density is shown on the right hand size of  Fig.~\ref{charge_dens} (please note again reversed ordering of hole levels) and behaves similarly to the ground electron state. 
The only difference is that $h_1$ seems to localize more strongly within the nanostructure, whereas $e_1$ shows some spreading (leakage) in the ($[110]$) direction perpendicular to the deformation. Generally all (10) hole states shown here are more strongly localized in the nanostructures area. This is expected since plots shown here must be consistent with numbers (degrees of localization reaching $96\%$) discussed earlier. 

Finally, let me note that the excited state $h_2$ shows a simple behaviour with a single nodal plane, but higher excited hole states have far more complicated nodal structure than electrons, with non-trivial nodes (both in $[1\underline{1}0]$ and $[110]$ directions) - an apparent manifest of holes multi-band character.

Apart from energies and charge densities, in the tight-binding formalism one can naturally inspect orbital contribution from different atomic orbitals constituting single particle quantum dot confined states. The orbital contribution to the ground hole state (for more details see the Appendix) is dominated by lateral $p$ atomic orbitals ($p_x$ and $p_y$) reaching $72-73\%$, with a significant fraction of $24\%$ from ($d_{yz}$ and $d_{zx}$) $d$-orbitals, and much smaller contributions from other states, most notably $p_z$, $d_{xy}$, and $s$ altogether constituting not more than about $3\%$ of the total hole ground state charge density.
For completeness, it should be mentioned that the ground electron state is built predominantly (over $90\%$ contribution) by $s$ and $s^*$ atomic orbitals with a small $4-5\%$ add-mixture of $p_z$ orbitals and much smaller contribution form other orbital species ($p_x$, $p_y$ and $d_{xy}$). 
Presence of $p$-type orbitals in the electron state or $s$ orbitals in the ground hole state should not surprise, and it is a direct manifestation of a multi-band and multi-valley character of the tight-binding calculation. Moreover, despite pronounced changes of single particle states energies with the deformation, the orbital contribution is very little affected and stable with the deformation (for more details please see Fig.~\ref{orbitalsandhhlh} in the Appendix).
Hence it can be speculated that the key features of elongated systems spectra are related to the overall, spatial character of single particle state, 
rather than their particular orbital components.

\begin{figure}
  \begin{center}
  \includegraphics[width=0.45\textwidth]{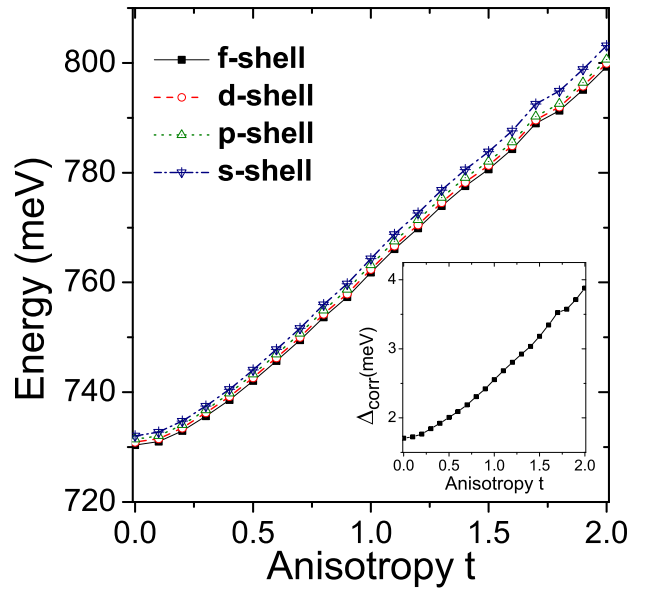}
  \end{center}
  \caption{Exciton ground state energy as a function nanostructure shape deformation \textit{t} along [1\underline{1}0] crystal axis.
   Lines correspond to cases with different number of single-particle shells ($s$,$p$,$d$ and $f$) included in the configuration interaction calculation. Inset: the difference between ground state X energy calculated using $s$-only and all shells up to $f$.}
  \label{Xenergy}
\end{figure}

Quasi-particle states are next used as the input for the many-body (up to $f$-shell) calculation of excitonic spectra and the resulting excitonic ground state energy evolution is shown on Fig.~\ref{Xenergy}.
Here within small deformations range the excitonic energy changes little since the decreased confinement in one direction ($[110]$) is compensated by the increased confinement in the other [1\underline{1}0] direction, consistent e.g. with my previous work on weakly elongated quantum dots~\cite{zielinski-elong} with $t\leq0.2$. The ground exciton state upward shift in energy is thus only $2.6$~meV for $t=0.2$ case with respect to $t=0$ system.
Yet, for larger deformations there is a clear, nearly linear growth of excitonic energy reaching about $800$~meV for $t=2.0$, that is about $69$~meV larger than for the non-elongated quantum dot.
It should be emphasized that the increase of excitonic ground state energy with an elongation ($69$~meV shift) is determined mostly by single particles energy contributions ($68$~meV) of electron and hole forming an exciton. There is thus apparently only a relatively small magnitude ($\approx 1$~meV) of change of electron-hole Coulomb interaction between cylindrical and elongated cases. However there are different contributions to electron-hole interaction due direct term and correction due to correlations - these effects will be discussed in the latter part of the text where excitonic complexes will be studied.

Apart from the direct electron-hole Coulomb interaction there is also a matter of configuration mixing.
Since in the single particle spectra we could observe the presence of different shells it is curious to check how these will affect the excitonic spectra.
Hence a series of configuration interaction calculations was performed with a systematically increasing number of electron and hole shells accounted for, and results summarized on Fig.~\ref{Xenergy}.
For the single exciton this corresponds to $4$, $36$, $144$, $400$ electron-hole configurations for $s$, $p$, $d$ and $f$ shells correspondingly.
Since there is no shell structure for elongated systems and moreover the $f$-shell is not well pronounced even for cylindrical dot, thus using a notion of ''shell'' is just a shortcut. Accounting for the $s$-shell means accounting for the ground electron and hole states only, $p$-shell corresponds to $3$ ($6$ with spin) lowest electron and $3$ ($6$ with spin) lowest hole states, and $d$-shell is $6$ ($12$ with spin) states for each of charged carriers, and finally $f$-shell corresponds to accounting for $10$ ($20$ with spin) of each of single particle states.
The energetic difference between cases accounting for $s$ and $f$ is shown as well (inset) on Fig.~\ref{Xenergy} and the correction due presence of higher shells is only about $1.7$~meV for $t=0$ symmetrical case, while it reaches more substantial $3.9$~meV for the largest considered deformation.
The trend shown on the inset on Fig.~\ref{Xenergy} is intuitive since one could expect higher role of configuration mixing in case of closely spaced levels for elongated systems of broken symmetry than in more cylindrical cases with well separated shells.
Whereas the effect of inclusion of higher-levels seems to be a small (few meV) correction as compared to the overall excitonic energy of about $750-800$~meV, yet
this particular ($s$-$f$) difference is a direct measure of correlation effects due to the add-mixture of higher energy configurations and will play a crucial role e.g. in the magnitude of excitonic complexes binding energies discussed later in the text.

\section{Excitonic fine structure: bright and dark excitons}

Apart from the main features it is curious to study the details of excitonic spectra, that is its fine structure.
Typical quantum dot spectra consist of two pairs of excitonic states.
One pair is formed by two states with anti-parallel electron-hole spin alignments leading to an optically active bright exciton doublet.
Second pair, energetically below the bright excitons (due to electron-hole exchange) is a doublet of two dark excitonic states with parallel electron-hole spins. Both of these doublets can be further split by various effects including presence of anisotropic lattice, strain, and alloying.~\cite{zielinski-elong}
The bright exciton splitting is also commonly known as the fine structure splitting.

\begin{figure}
  \begin{center}
  \includegraphics[width=0.45\textwidth]{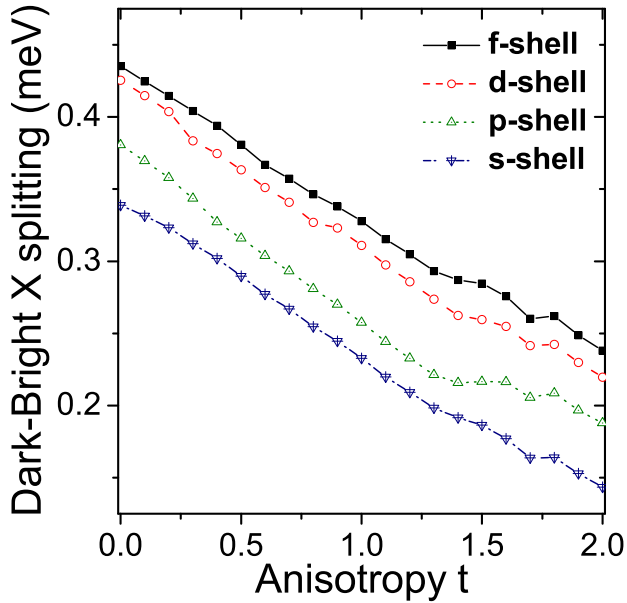}
  \end{center}
  \caption{The bright-dark exciton splitting (electron-hole exchange splitting) as a function of nanostructure shape deformation \textit{t} along [1\underline{1}0] crystal axis.
  Lines correspond to cases with different number of single-particle shells ($s$,$p$,$d$ and $f$) included in the configuration interaction calculation.}
  \label{DB}
\end{figure}

Fig.~\ref{DB} shows energy difference between the lowest bright exciton state and the higher energy dark exciton.
Notably this splitting goes down with the elongation in all considered approximations in the configuration interaction calculation.
We can understand that in terms of decreasing electron-hole exchange interaction due to change of lateral confinement with a progressing deformation, however we should also note that this energy difference is affected also by splittings within bright and dark doublets discussed later.
Interestingly one can also observe here a non-negligible effect of higher-lying levels on dark-bright exciton energy splitting.
Here, the calculation including the $s$-shell only gives the splitting by about $100$~meV smaller than that performed in the full $f$-shell basis.
The difference between the $d$-shell and the $f$-shell cases is much smaller and varies between $10$ to $15$~meV, and is somewhat larger for elongated cases.

One could question here the convergence of configuration interaction method with respect to number of shells included.
In fact I will try to address this problem in a future work, however currently accounting for shells even higher than $f$ is prohibitively numerically demanding.
For the dark-bright exciton splitting a rough estimate of such error would be about $30$~$\mu$eV,
still much smaller than the absolute value of the dark-bright exchange splitting.

\begin{figure}
  \begin{center}
  \includegraphics[width=0.45\textwidth]{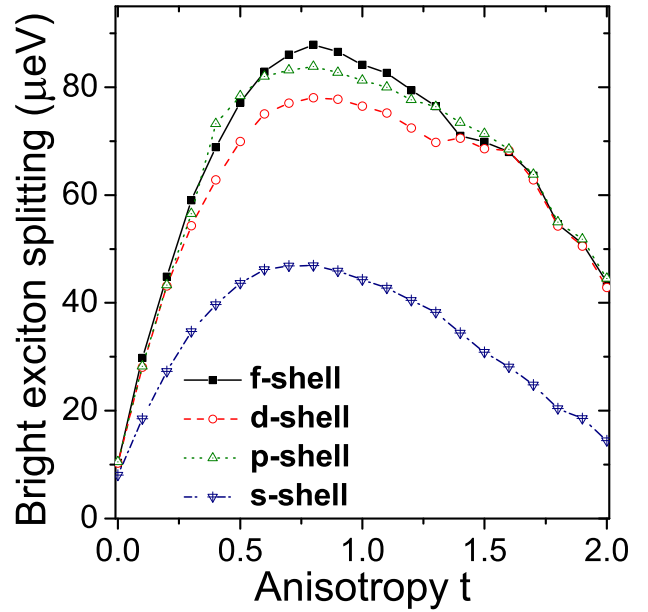}
  \end{center}
  \caption{The bright exciton splitting (anisotropic electron-hole exchange splitting) as a function of deformation \textit{t}.
   Lines correspond to cases with different number of single-particle shells ($s$,$p$,$d$ and $f$) included in the configuration interaction calculation.}
  \label{BES}
\end{figure}

Next, Fig.~\ref{BES} shows the splitting of the bright exciton (the fine structure splitting) as a function of the deformation.
One observes here two characteristic and peculiar features.
Firstly, there is again a significant difference between configuration interaction results when accounting for different shells.
A ''customary'' approach with $s$-shell only reproduces well the general trend, but quite importantly it strongly underestimates the magnitude of the splitting giving about half of it. Addition of higher energy configurations has little effect for weakly deformed systems, e.g. for a cylindrical case (where the fine structure splitting is present due to low lattice symmetry, despite the cylindrical shape) the contribution from the $s$-shell equals to $8$~$\mu$V and is a dominating the total splitting equal to $10.5$~$\mu$V. However the importance of higher levels is pronounced for elongated systems and it can be intuitively understood since the reduced symmetry would seem to promote increased mixing between configurations.
For a moderately deformed nanostructures and $t$ parameter value between about $0.5$ and $1.2$ the fine structure splitting dependence on the number of included shells is far from trivial, e.g. the addition of $d$-shell apparently reduces the bright exciton splitting compared to $p$-shell inclusion only.
Interestingly for larger deformations ($t>1.5$) addition of higher shells (above the $p$-shell) has a small effect on the splitting.
For the largest considered deformation the contribution due to $s$-shell only gives splitting of $14.4$~$\mu$eV, addition of the $p$-shell increases the splitting by additional $28.5$~$\mu$eV giving the total splitting of $44.5$~$\mu$eV, further addition of $d$- and $f-$ curiously decreases the splitting, yet by mere $0.7$~$\mu$eV.
It thus appears that for elongated cases most of the splitting is given by accounting for the ground, first and second excited electron and hole levels whereas the ground state orbitals give as little as $1/3$ of the contribution to the splitting.

Moreover our previous results demonstrate the that bright exciton splitting is also a sensitive function of overall nanostructure volume (height and length) as well as intermixing effects due to alloying in the surrounding barrier. 

Apart from observations regarding the correlations due to different shells the next key effect is the trend of fine structure evolution itself.
At first the splitting goes up quasi-linearly with $t$ following predictions based on the effective mass approximation~\cite{kadantsev-eh} and anisotropy related mixing of quantum dot heavy-hole states.
However with the increase of the deformation the trend seems to flatten-up, and it saturates with a plateau ($\approx 80$~$\mu$eV) at about $t$ equal to $0.7-0.8$.
More curiously with further shape deformation the fine structure splitting on Fig.~\ref{BES} is reduced with anisotropy: a trend clearly contradicting an simple intuition.

\begin{figure}
  \begin{center}
  \includegraphics[width=0.45\textwidth]{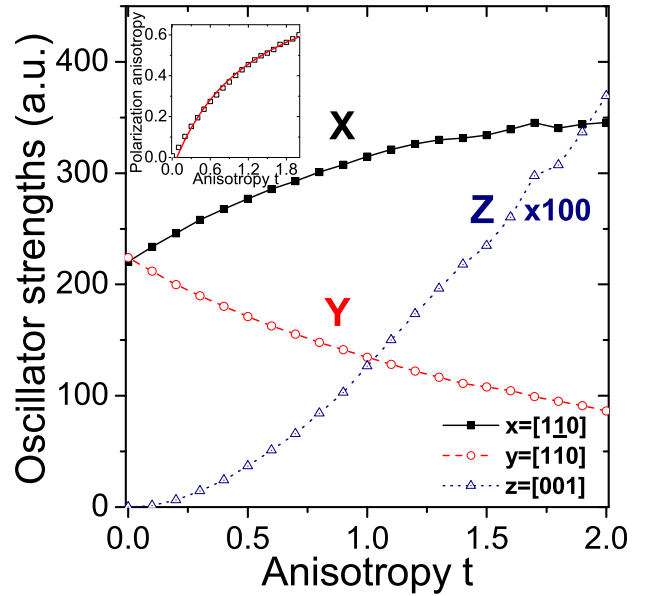}
  \end{center}
  \caption{The calculated excitonic optical spectrum (oscillator strengths) for the bright excitons (in-plane polarizations) and the dark exciton (out-of-plane, "z" polarization)
  as a function of nanostructure deformation \textit{t}.
  The inset shows bright exciton polarization anisotropy.}
  \label{intensity}
\end{figure}

In order to analyze this effect further let us focus our attention on the excitonic emission spectra.
Fig.~\ref{intensity} shows emission from two bright excitonic states in form of two $X=[1\underline{1}0]$ and $Y=[110]$ linearly polarized lines.
$X$ line corresponds to lower energy bright exciton, whereas $Y$ to higher energy bright exciton state.
For $t=0$ case there is only a small intensity difference between these two lines due to the presence of anisotropic crystal lattice and strain.~\cite{zielinski-elong}
For $t>0$ the dominant oscillator strength comes from $X$ polarization, following thus an elongation axis, whereas for the $Y$ polarized line the oscillator strength is reduced with
elongation, with the ratio $I_{max}/I_{min}=I_{X}/I_{Y}$ going from $\approx1$ for $t=0.0$ to over $4$ for $t=2.0$.
The growth of $X$-polarized line oscillator strength is thus strictly followed by the decrease in $Y$-polarized line oscillator strength.
Curiously the $I_{max}/I_{min}$ dependence on elongation can be approximately described by a following linear relation $I_{max}/I_{min}\approx 1+3/2*t$.
A more established measure of polarization anisotropy is polarization degree $C=\left(I_{max}-I_{min}\right)/\left(I_{max}+I_{min}\right)$ shown (black/empty squares) on the inset on Fig.~\ref{intensity}.
Polarization degree dependence on elongation ratio can be well described by a simple following relation: $C=t/(4/3+t)$.
A more accurate formula can be found by a fitting with $C=(t-0.0725)/(1.26544+$t), shown as a red line on the inset on Fig.~\ref{intensity}.
Such a simple relation between polarization anisotropy and elongation suggest that there maybe a single dominant factor strongly affecting the spectra.

It is often assumed~\cite{bayer-eh} that structural quantum dot elongation will lead to mixing of the bright exciton states
$\left|1\right>=\left|\frac{3}{2},-\frac{1}{2}\right>$ and $\left|-1\right>=\left|-\frac{3}{2},\frac{1}{2}\right>$, where
$\pm\frac{3}{2}$ corresponds to heavy-hole spin projection, and $\pm\frac{1}{2}$ are electron spin projections.
Such mixing would lead to the bright exciton splitting and presence of two orthogonal lines of linear polarization.
However mixing of pure heavy-hole states seem not to be sufficient to describe strong polarization anisotropy as observed here.
Recently several authors~\cite{Smolenski.PRB.2012,tonin,Tsitsishvili,leger} have suggested a mechanism in which lateral anisotropy would effectively induce heavy-hole/light-mixing,
with the bright excitons states given as:
\begin{equation}
\left|\pm \widetilde{1}\right>=\sqrt{1-\beta^2}\left|\pm\frac{3}{2},\mp\frac{1}{2}\right>+\beta\left|\pm\frac{1}{2},\mp\frac{1}{2}\right>
\end{equation}
thus effectively add-mixing exciton state heavy-hole component $\left|\frac{3}{2}\right>$ with a light-hole component of opposite projection $\left|-\frac{1}{2}\right>$, and correspondingly mixing $\left|-\frac{3}{2}\right>$ with $\left|\frac{1}{2}\right>$.
$\beta$ is a measure of that mixing, and the polarization anisotropy due to mixing can be descried as:\cite{tonin,leger}
{\begin{equation}
C\left(\beta\right)=\frac{2\beta\sqrt{3\left(1-\beta^2\right)}}{3-2\beta^2}
\label{deg-pol}
\end{equation}}

We can use a fit to the above equation in order to effectively retrieve $\beta$ from our atomistic calculations of the polarization degree.
\begin{figure}
  \begin{center}
  \includegraphics[width=0.45\textwidth]{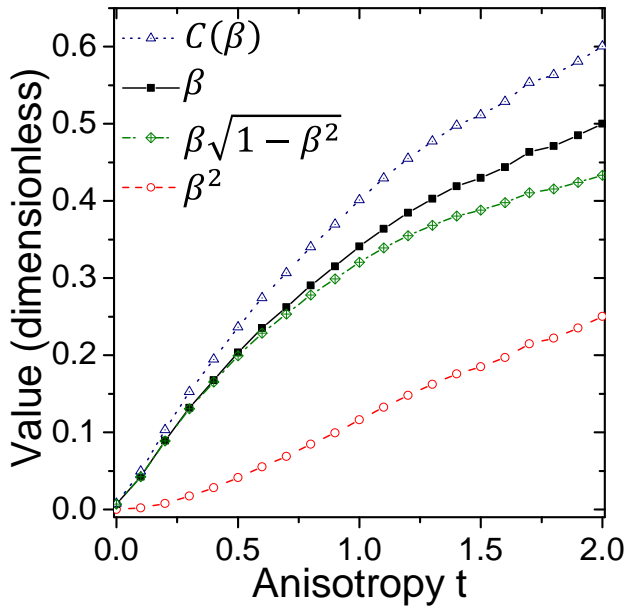}
  \end{center}
  \caption{The bright exciton polarization anisotropy $C\left(\beta\right)$ (same as on the inset on Fig.~\ref{intensity}) compared with
   strength of light-hole/heavy-hole mixing $\beta$ as well as different functions of $\beta$ as functions of nanostructure deformation \textit{t}.}
  \label{beta}
\end{figure}
The results are shown on Fig.~\ref{beta}, with $\beta$ increasing monotonically with the deformation $t$ and reaching considerable value of about $\approx 0.5$ for the largest considered deformation. My previous~\cite{zielinski-tallqd} calculations for cylindrical nanowire quantum dots indicate that further increase of aspect ratio could in fact lead to a formation of a light-hole dominated exciton, however the analysis of light-hole excitons goes beyond the scope of current work.
Additionally and contrary to tall (high vertical aspect ratio) nanowire quantum dots~\cite{zielinski-tallqd} the orbital contribution of single particle states in flat ($3$~nm of height) nanostructure is not changed much with the deformation (for the ground hole state in particular; see the discussion earlier and the Appendix) and the large $\beta$ content corresponds to a change in the envelope character of the hole wave-function, rather than in its microscopic part. Here the term ''envelope'' was borrowed from the language of the continuous media approximation (and methods such as the effective mass approach) and it should be used with a great caution since the notion of ''envelope'' is not present in the linear combination of atomic orbitals (LCAO) approach utilized by the tight-binding method.

\begin{figure}
  \begin{center}
  \includegraphics[width=0.45\textwidth]{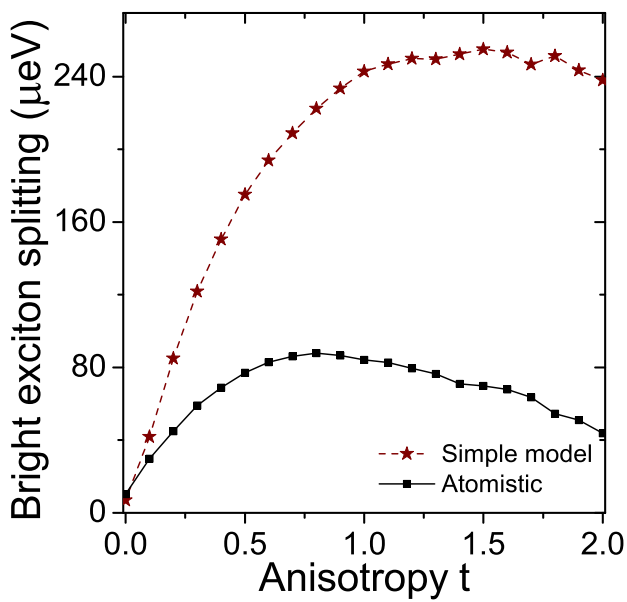}
  \end{center}
  \caption{Estimation of the bright exciton splitting from a phenomenological model and from an atomistic calculations
   as a function of nanostructure deformation \textit{t}.}
  \label{simplemodel}
\end{figure}

Let us now come back to the matter of the bright exciton spectra as discussed above and shown earlier on Fig.~\ref{BES}.
For clarity we plot this splitting (for $f$-shell case only; solid-line/squares) again on Fig.~\ref{simplemodel}, yet this time compared with simple model estimation (dashed line/starts) based on recent work by Tsitsishvili~\cite{Tsitsishvili}, where the bright exciton splitting in case of significant light-hole add-mixture could be given $\approx \frac{4}{\sqrt{3}}\beta\sqrt{1-\beta^2}\Delta_{ST}$, where $\Delta_{ST}$ is the ''usual'' (i.e. non-anisotropic) electron-hole exchange, related to dark-bright exciton splitting shown previously on Fig.~\ref{DB}.
The dependence $\beta\sqrt{1-\beta^2}$ term is shown on Fig.~\ref{beta}, where it is compared with $\beta$ and $\beta^2$ and it shows sub-linear dependence on $t$.
This term is then multiplied by the dark-bright exciton splitting $\Delta_{ST}$, which I reiterate gets reduced with the increasing deformation (Fig.~\ref{DB}).
Hence for elongated systems there are two opposing effects.
The first is the increasing ($\beta\sqrt{1-\beta^2}$) contribution due to light-hole add-mixture that leads to the increase of the splitting.
The second opposing effect is the decreasing value of the electron-hole overlap leading to a reduction of the dark-bright exciton splitting.
These two effects combined together lead first to the growth of the bright exciton splitting with the deformation,
then its saturation and then the decrease of the splitting with even further elongation.
The simple model, combined with input from the atomistic calculation ($\Delta_{ST}$ and $\beta$ from excitonic energy and emission spectra correspondingly) is thus able to qualitatively explain the behaviour of the bright exciton in highly-elongated system, yet we should also emphasize it is only a qualitative agreement.
The phenomenological model overestimates the value of splitting by a large factor (typically over $3$) and the fully atomistic calculation is needed for the qualitative prediction of the splitting magnitude.

Based on the above one can conclude that the strong admixture of light-hole component could explain polarization properties of the bright excitons and their splitting in strongly elongated systems, moreover it turns out that light-excitons would also have a pronounced effect on the dark exciton proprieties.
My calculations indicate that with an increasing deformation one of the dark excitons gets a non-negligible oscillator strength, shown on Fig.~\ref{intensity} as $z$ (out-of-plane) polarized line, with magnitude reaching considerable (about $1\%$) fraction of the $X$-polarized bright exciton emission.
This dark exciton non-zero optical activity comes from higher energy dark exciton state, where the other (low energy) state remains fully dark, consistent with group-theoretical predictions for heavy-hole dominant $C_{2v}$ quantum dots.~\cite{Dupertuis.PRL.2011,karlsson}
It should be noted however that such large dark exciton optical activity is orders of magnitude stronger than that observed for weakly deformed quantum dots.~\cite{Korkusinski.PRB.2013,zielinski-elong}
In a phenomenological model accounting for the light-hole exciton add-mixture to the dark exciton state,~\cite{Smolenski.PRB.2012,Tsitsishvili} the magnitude of the dark exciton activity should~\cite{Smolenski.PRB.2012} follow $\beta^2$ dependence.
This is in fact qualitatively consistent with our results: with $z$-polarized line on Fig.~\ref{intensity} and $\beta^2$ on Fig.~\ref{beta} showing qualitatively the same behaviour.
We note here that such pronounced optical activity of the dark of exciton could have potential applications in quantum information and computation~\cite{PhysRevB.92.201201,PhysRevX.5.011009,Zielinski.PRB.2015} since the dark state is both long-lived and optically addressable at the same time, and
its properties could be likely be tailored by the degree of nanostructure elongation.

\begin{figure}
  \begin{center}
  \includegraphics[width=0.45\textwidth]{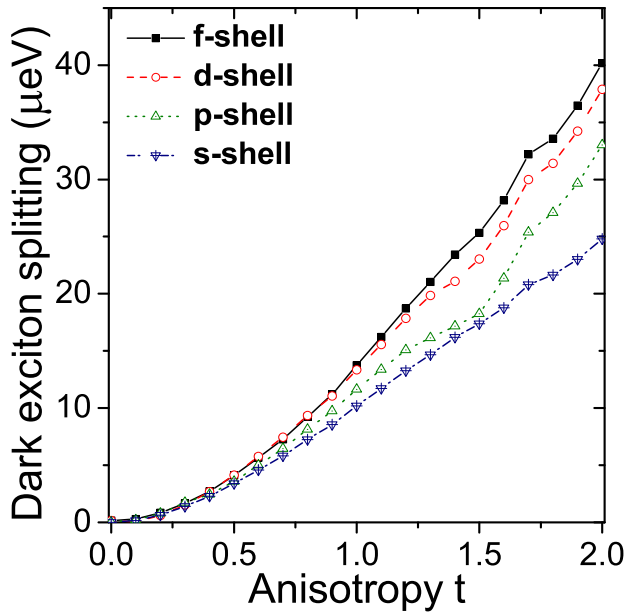}
  \end{center}
  \caption{The dark exciton splitting as a function of deformation \textit{t}.
   Lines correspond to cases with different number of single-particle shells ($s$,$p$,$d$ and $f$) included in the configuration interaction calculation.}
  \label{DES}
\end{figure}

Apart from the optical activity, it is instructive to study the dark exciton splitting as shown on Fig.~\ref{DES}.
The dark exciton splitting goes from a small value of $0.14$~$\mu$eV for a cylindrical $t=0$ system to a substantial splitting exceeding $40$~$\mu$eV for $t=2.0$.
One should notice that this is an extraordinarily large dark exciton splitting, comparable with that of the bright exciton.
Based on a simple model~\cite{Tsitsishvili} one can expect the dark exciton splitting to grow proportionally to $\beta$ (and thus to $t$),
due the admixture of the light-hole exciton add-mixture to the dark exciton $\left|\pm 2\right>$ state.
Therefore a substantial content of the light-hole exciton will have a dramatic impact on the dark exciton properties, including the dark exciton splitting.
One can observe however yet another effect that cannot be easily captured even qualitatively by a simple model.
Namely, for deformations $t$ larger than $0.5$ exciton one can notice the increasing role of higher-lying shells,
with the magnitude of dark exciton splitting increasing progressively with the admixture of these shells.
This leads to a general trend of the dark exciton emission resemble more quadratic-like behaviour as a function of $t$ with $f$-shell included, and more linear-like for the $s$-shell included only.

To summarize this section let me note that the calculated range (from $10$ to $88$~$\mu$eV) of the bright exciton splittings falls reasonably well within the range of values observed experimentally~\cite{dash-mrowinski} for large-aspect ratio quantum dashes emitting at around $800$~meV (i.e. from about $25$ to nearly $200$~$\mu$eV). However I emphasize that in this work I do not aim at the direct comparison with a particular quantum dash or quantum dot experiment, but rather focus on an analysis of general trends with shape elongation. 
Moreover a detailed comparison with a particular experiment~\cite{dash-mrowinski} should include a 	comprehensive knowledge of nanostructure's morphology, which is often near impossible to obtain.~\cite{dery2004nature,sauerwald2005size,khan2014self,Mlinar-inverse}
Nevertheless one can speculate here that obtained results strongly suggest that is is unlikely for $[1\underline{1}0]$ elongated quantum dashes to achieve a very small fine structure splitting unless other phenomena such as alloying due to annealing or composition intermixing are included into the consideration~\cite{olbrich2017polarization,jung2018effect} or external fields are applied.~\cite{mrowinski2015magnetic}
Otherwise growers should aim for non-elongated~\cite{kors2018telecom} or weakly $[110]$ elongated nanostructures, where small elongation along $[110]$ axis~\cite{zielinski-elong} should lead to the reduced fine structure splitting, by minimizing the anisotropic contribution due to strain.
One the other hand the profoundly increased dark exciton states optical activity in highly-elongated systems could in principle open new routes for quantum dashes applications including entanglement generation via the time-bin approach~\cite{PhysRevLett.94.030502,weihs2017time} or by utilizing the dark exciton as a optically addressable quantum bit.~\cite{Poem.Nature.2010,PhysRevB.92.201201,PhysRevX.5.011009,Zielinski.PRB.2015}

\section{Excitonic complexes}
Besides a single exciton is it instructive to study spectral properties of excitonic complexes, i.e. positively ($X^+$),
and negatively ($X^-$) charged excitons, and the biexciton ($XX$).
Understanding of the biexciton spectra is particularly relevant from applications point of view since in various schemes~\cite{PhysRevLett.84.2513,weihs2017time} 
its plays an essential role for the entanglement generation.

\begin{figure}
  \begin{center}
  \includegraphics[width=0.45\textwidth]{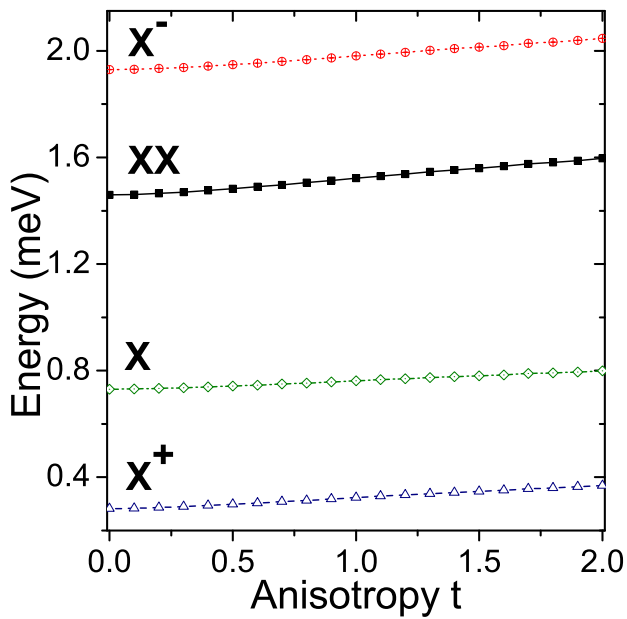}
  \end{center}
  \caption{Ground states energy evolution for the biexciton ($XX$), exciton ($X$), positively ($X^+$) and negatively ($X^-$) charged trions as a function of nanostructure deformation \textit{t}.}
  \label{complexes}
\end{figure}

\begin{figure}
  \begin{center}
  \includegraphics[width=0.45\textwidth]{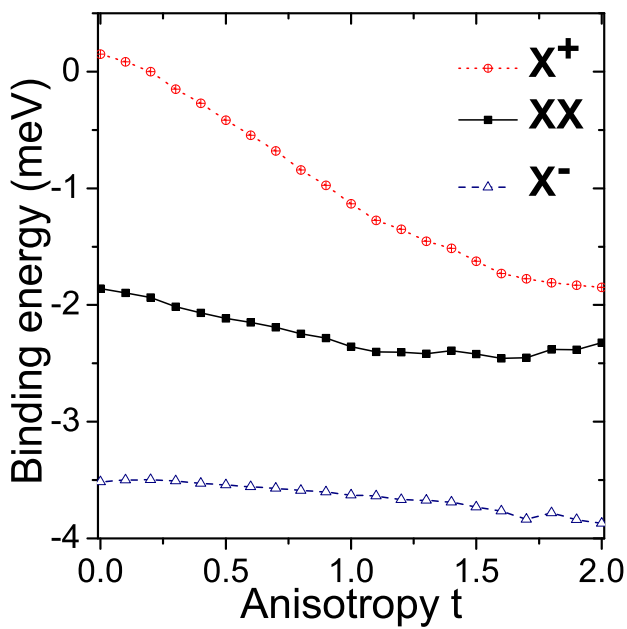}
  \end{center}
  \caption{Binding energies of the biexciton ($XX$), positively ($X^+$) and negatively ($X^-$) charged trions as a function of nanostructure deformation \textit{t}.}
  \label{bindings}
\end{figure}

The ground states energies of these complexes are presented on Fig.~\ref{complexes}, and they show a weak monotonic, increasing trend with $t$, similar to 
that of the single exciton ($X$) shown previously on Fig.~\ref{Xenergy}, and presented here as well for the purpose of comparison.
Contrary to a single exciton, in an optical experiment it is not straightforward to measure directly ground state energies, but one rather studies transition energies.
Therefore Fig.~\ref{bindings} shows the evolution of binding energies of $X^+$, $X^-$, and $XX$ as a function of shape deformation $t$.
Similarly to a photoluminescence experiment these binding energies were calculated with respect to the single exciton.
For a case of $XX$ this means the optical recombination leaving $X$ as a final state.
The energy of this bright optical transition (i.e. $E_{XX}-E_{X}$) is then compared with that of the bright exciton $E_{X}$ to define the binding energy of a complex.
The binding energy measures thus the energetic difference between bright 
lines from recombining $XX$, $X^+$, $X^-$ complexes and the lowest bright $X$ state.

Depending on the nature of a complex, their binding energy spectra reveal a different behaviour, with the $XX$ and $X^-$ showing
only a weak change with the shape elongation.
The $XX$ binding energy varies from about $-1.9$ to about $-2.5$~meV, similar to our earlier 
results~\cite{dash-mrowinski} obtained for quantum dashes with triangular cross-sections.
Whereas the $X^-$ is more bound with the binding energy changing from $-3.5$ to nearly $-3.9$~meV for the largest considered deformation.
Here I utilize a convention in which a negatively bound complex has emission energy lower than the bright single exciton state.

The positively charged exciton $X^+$ shows a much more pronounced trend with the shape elongation.
Its binding energy goes form a small positive value of $0.15$~meV for a cylindrical system to nearly about $-1.8$~meV for $t=2.0$.
It is curious to analyze such different behavior of $X^+$, and since $XX$ and $X^+$ differ by a presence of an additional hole, one can speculate
that the difference between all these excitonic species is related to the hole state properties.

On a more formal ground one can estimate these binding energies in terms of Coulomb integrals calculated for an electron and a hole in their ground states $e_1$ and $h_1$:~\cite{gong-prb77,Zielinski-natural,ZielinskiSubstrate2012}

\begin{align}
\Delta E(XX) & = J_{e1e1}+J_{h1h1}-2J_{e1h1}+\Delta_{corr}(XX-X) \nonumber \\
\Delta E(X^-)& = J_{e1e1}-J_{e1h1}+\Delta_{corr}(X^--e) \nonumber \\
\Delta E(X^+)& = J_{h1h1}-J_{e1h1}+\Delta_{corr}(X^+-h)
\label{estimation}
\end{align}
where $J$ are electron-electron, hole-hole, and electron-hole Coulomb integrals mentioned above, $\Delta_{corr}$ are (negative) corrections due to correlation
(and exchange) effects, and accounted for by the exact diagonalization (configuration interaction).
Similarly to experiments~\cite{dash-mrowinski} the recombination process happens between the initial (e.g. $XX$) and a final (e.g. $X$) state of two different
excitonic complexes.

\begin{figure}
  \begin{center}
  \includegraphics[width=0.45\textwidth]{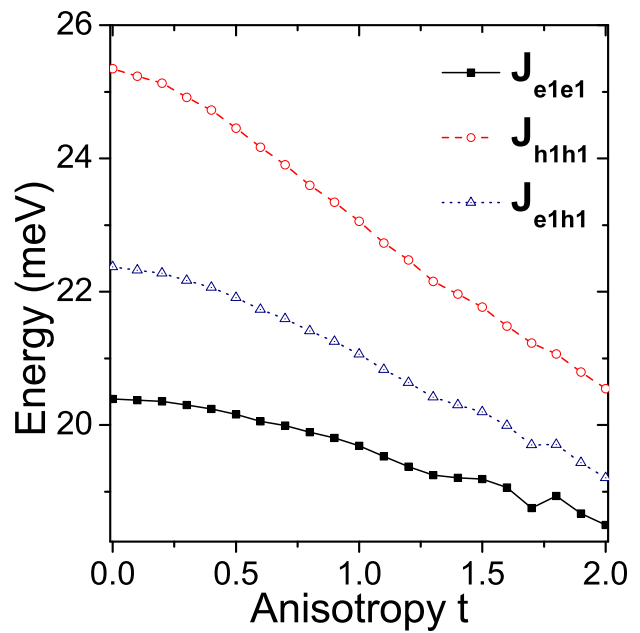}
  \end{center}
  \caption{Electron-electron ($J_{e1e1}$), electron-hole ($J_{e1h1}$) and hole-hole ($J_{h1h1}$) direct Coulomb integrals calculated for electron and hole
   occupying their ground states ($e_1$ and $h_1$) as a function of nanostructure deformation \textit{t}. Please note the same sign convention used for all these terms.}
  \label{coulombs}
\end{figure}

The above formula should in principle help us to understand properties of excitonic complexes in terms of contributions from selected Coulomb integrals, which are shown on Fig.~\ref{coulombs}.

The hole-hole repulsion is reduced by about $5$~meV with the anisotropy, whereas the electron-electron repulsion is reduced by a $1.9$~meV factor only,
and the electron-hole attraction shows an intermediate trend (sign of $J_{e1h1}$ was set as positive according to convention used in Eq.~\ref{estimation}).
The pronounced change of an integral involving hole state suggests that the hole is apparently more prone to the reduction
of lateral confinement (shrinking) in the $[110]$ direction, perpendicular to the shape elongation in $[1\underline{1}0]$.
Although likely oversimplified this is consistent with charge density plots shown earlier on Fig.~\ref{charge_dens}, where $h_1$ seems to be more localized in the nanostructure for elongated cases, whereas $e_1$ tends to leak out, thus partially overcoming the effect of elongation.

\begin{figure}
  \begin{center}
  \includegraphics[width=0.45\textwidth]{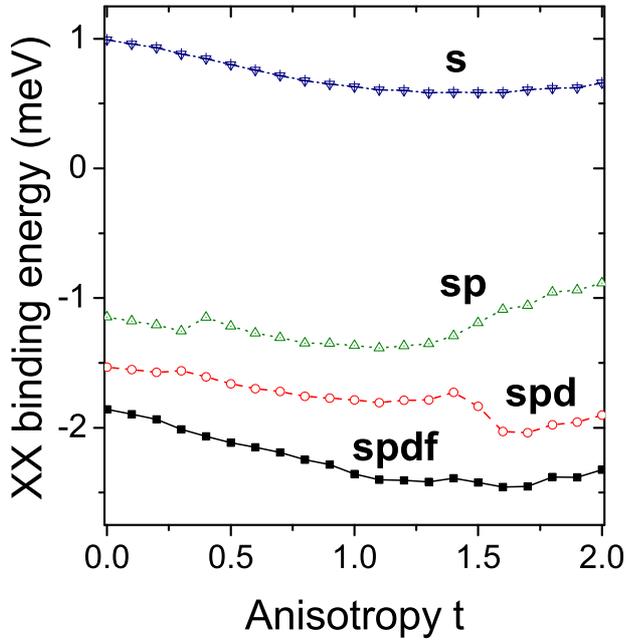}
  \end{center}
  \caption{Evolution of the biexciton binding energy as a function of nanostructure deformation \textit{t} and the number of single particle shells included in the
   configuration interaction calculation.}
  \label{XXconv}
\end{figure}

The strongly decreasing value of $J_{h1h1}$ integral combined with $\Delta E(X^+) = J_{h1h1}-J_{e1h1}$ formula (from Eq.~\ref{estimation})
seems to be able to qualitatively explain the strong $X^+$ evolution with the deformation, and also weak changes of $X^-$ and $XX$ states.
However the simple estimations based solely on the ground state properties are not able to address the overall binding energy of excitonic complexes.
This is well illustrated on Fig.~\ref{XXconv}, where the $XX$ binding energy was calculated by using different number of electron and hole levels accounted for in the configuration interaction calculation. The $s$-shell calculation effectively corresponds to Eq.~\ref{estimation} with correction due to correlations $\Delta_{corr}(XX-X)$ set to zero.
Whereas the overall trend with the deformation does not change strongly between different approaches, only inclusion of $s$,$p$,$d$ and $f$-shells seems to produce binding
energy reaching $-2.5$~meV in a reasonable agreement with the experiment.~\cite{dash-mrowinski,mrowinski-pol}

\begin{figure}
  \begin{center}
  \includegraphics[width=0.45\textwidth]{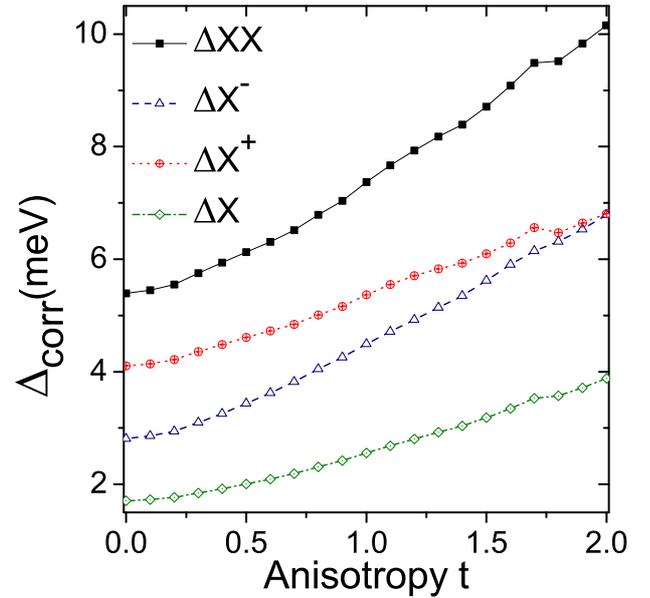}
  \end{center}
  \caption{Correction due to configurations mixing for the biexciton ($XX$), exciton ($X$), positively ($X^+$) and negatively ($X^-$) charged trions as a function of nanostructure deformation \textit{t}.}
  \label{corr}
\end{figure}

For the case including the $f$-shell, the correction to $XX$ binding energy due to correlations $\Delta_{corr}(XX-X)$ is equal to about $-3$~meV for all considered $t$ values.
However it should be emphasized again that the correction due to correlations ($\Delta_{corr}(XX-X)$) in the $XX$ binding energy in the recombination process is affected by both the correlations
in the initial $XX$ state ($\Delta_{corr}(XX)$) and the final $X$ states ($\Delta_{corr}(X)$).
Correlations in the initial state, i.e. in the $XX$ ground state (Fig.~\ref{corr}), shift down its energy by $5.4$~meV for $t=0$ as compared to a single electron-hole $s$-shell configuration. This correction grows with the deformation and reaches a substantial $10$~meV for $t=2.0$. A pronounced ground $XX$ state energy shift due to configuration mixing is also well known in standard, non-elongated quantum dots~\cite{arek-book}, and is a key factor determining $XX$ binding energy in the emission spectra.
Yet the magnitude of this correction reaching $10$~meV is far larger than in typical self-assembled quantum dots.
Moreover here differently from cylindrical systems the correlation correction is also present and non-negligible in a final, single exciton state. 
This effect was briefly mentioned in the earlier part of this paper (inset on Fig.~\ref{Xenergy}), with correlation correction reaching considerable value of $-3.9$~meV for a single exciton in its ground state and $t=2.0$. In elongated systems correlations thus have a strong effect on both (initial) $XX$ and (final) $X$ states, and in both these of excitonic species, these correction increase their magnitudes with the deformation. Yet effectively in $XX$ recombination spectra case these terms partially cancel out, leading to apparently stable biexciton binding energy with the elongation. Finally, I speculate the that addition of higher excited electron and hole levels (currently prohibited by numerical complexity) would likely further reduce the $XX$ binding energy closer to the experimentally observable values, i.e. about $-3$~meV.

\section{Summary}
An increasing nanostructure shape anisotropy leads to transition from a cylindrical quantum dot to an heavily-elongated quantum dash.
In this work this transition was shown to have a strong impact on both the single particle and many-body properties of such systems.
As expected, the shape deformation leads to an increasing splitting of the bright exciton, however at a certain aspect ratio the splitting saturates and peculiarly it is reduced with the further elongation. The emission from the bright exciton is characterized as well by a strong degree of polarization anisotropy.
The magnitudes of fine structure splitting ($40-90$~$\mu$eV) as well the polarization degree of up to $0.6$, obtained here by atomistic calculations, compare well with experimental data for InAs/InP quantum dashes, even though we did not focus on alloying (mixed composition) effects always present in epitaxial systems. The main spectral features of strongly elongated systems can be also qualitatively explained in terms of a simple model assuming anisotropy induced contribution of a light-hole exciton to a heavy-hole dominated excitonic ground state. This contribution affects the dark excitonic states as well, leading to a dark exciton splitting growing proportionally with the shape deformation, and the magnitude of splitting reaching comparable values to that of the bright exciton.
Moreover the dark exciton gains a very significant increase of its optical activity getting to about $1\%$ of the dominant bright exciton line.
The shape elongation was further found to have a strong impact on correlation effects in these nanostructures.
The add-mixture of higher, excited (often with tails in the wetting layer) electron and hole states (up to the $10$-th level corresponding the $f$-shell of a cylindrical quantum dot) has a pronounced effect on the ground excitonic complexes states, 
magnitudes of dark and bright excitons splitting of the single exciton, as well as binding energies of excitonic complexes.
Finally, this paper shows that the control of the degree of elongation (or quantum dot anisotropy) could in principle be used to tailor optical properties of nanostructures in a broad range of values, including cases with a curiously strong optically activity of the dark exciton.

\section{Acknowledgment}
The author would like to thank Michał Gawełczyk, Krzysztof Gawarecki, and Grzegorz Sęk for insightful discussions. The  support from the Polish National Science Centre based on decision No. 2015/18/E/ST3/00583 is kindly acknowledged.

\section{Appendix}
\begin{figure}[h]
\begin{tabular}{ll}
\includegraphics[scale=0.22]{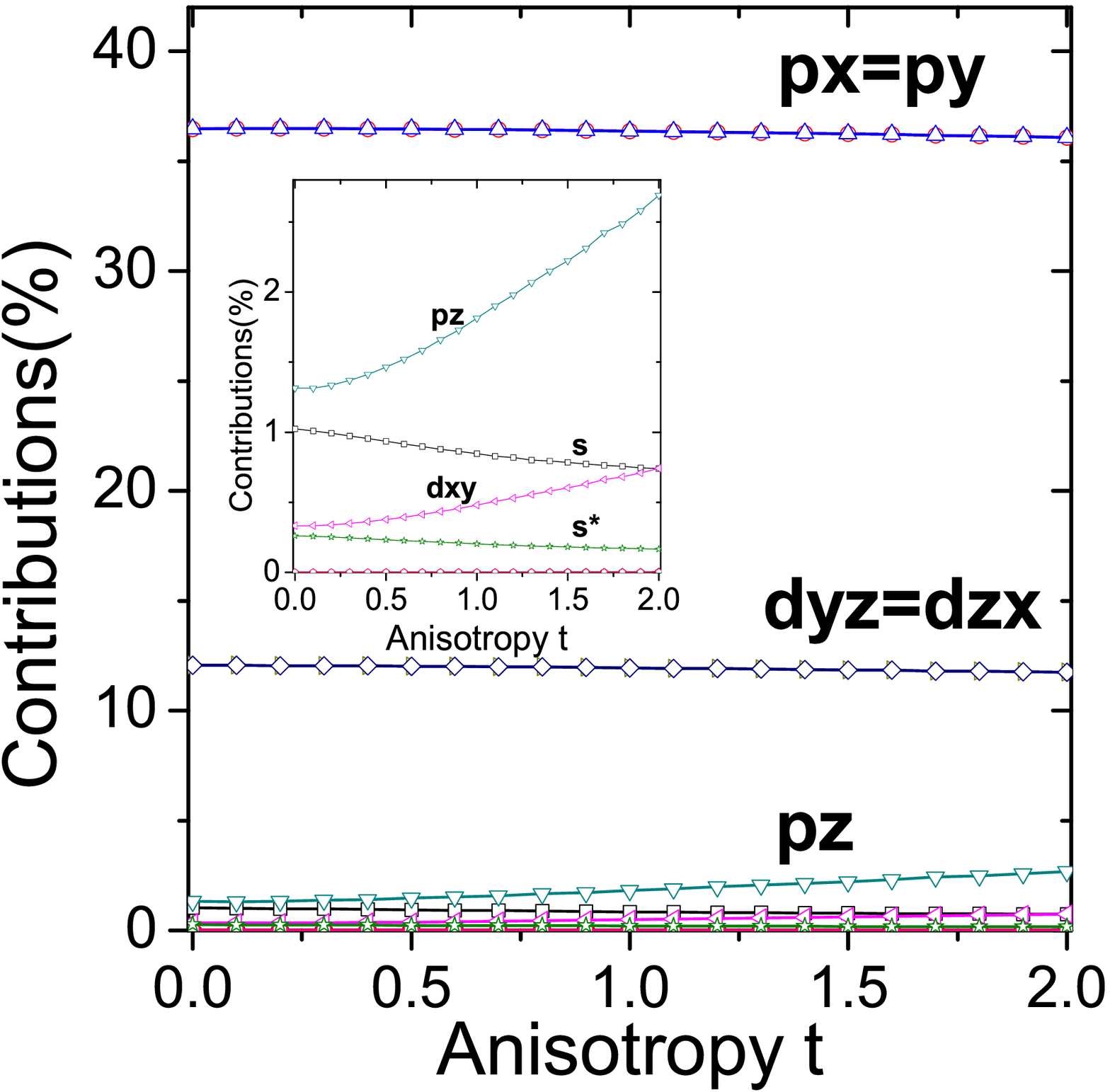}
&
\includegraphics[scale=0.22]{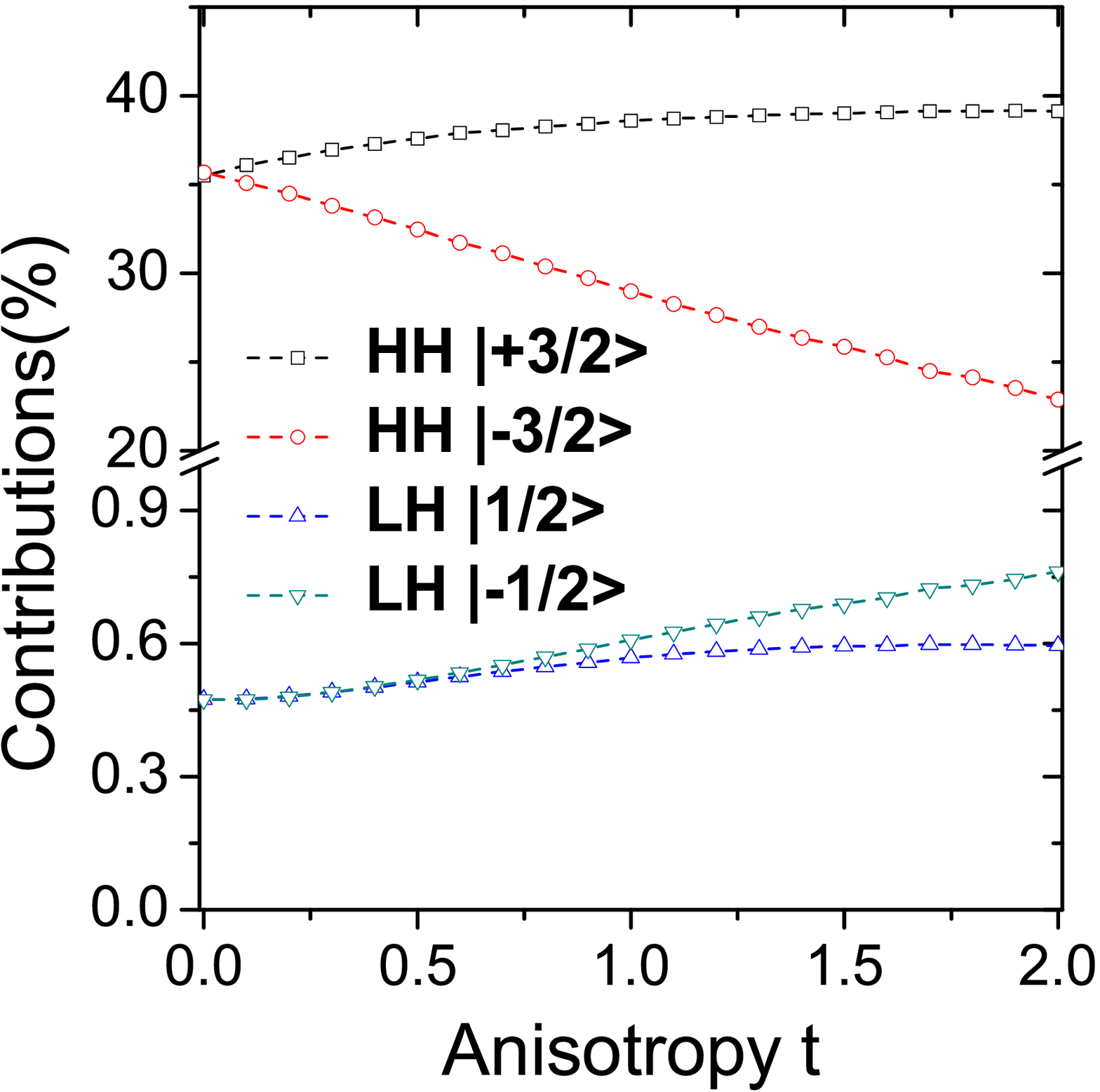}
\end{tabular}
\caption{Atomic orbital contributions (left) and combination of atomic orbital constituting light- (LH) and heavy- (HH) hole-like combinations as a function of nanostructure deformation \textit{t}.}
\label{orbitalsandhhlh}
\end{figure}

Since the ground hole single particle state has a dominant contribution to several lowest (two dark and two bright) excitons state and due to the importance of heavy-hole/light-hole couplings as discussed above let us inspect atomic orbital coefficients in the tight-binding (LCAO - linear combination of atomic orbitals) expansion of the ground hole state wave-function shown on Fig.~\ref{orbitalsandhhlh} (left).
One should analyze such plots as Fig.~\ref{orbitalsandhhlh} with care and note first that apart from atomic orbital contribution there is an strong ''envelope'' contribution which is not straightforward to be retrieved from the tight-binding method, and is not shown on Fig.~\ref{orbitalsandhhlh}.
Either way, inspection of Fig.~\ref{orbitalsandhhlh} reveals that (sum of squared moduli) contributions from $p_x$ and $p_y$ orbitals are the same and only weakly depend on the deformations. Combined contribution of lateral $p$ orbitals reaches $72-73\%$ and orbital-wise is a dominant contribution to the wave-function.
Moreover one can notice important contribution of about $24\%$ combined from $d_{yz}$ and $d_{zx}$ orbital, which again does not change with the deformation.
The elongation however affects $p_z$ orbitals contribution which goes from about $1.3\%$ for $t=0$ to almost $2.7\%$ for $t=2.0$.
Contributions from $s$ (and $s^*$) are (not surprisingly) much smaller and do not exceed $1\%$ (inset on Fig.~\ref{orbitalsandhhlh} (left)).
Here $z$ corresponds to $[001]$ direction and it must be strongly emphasized that $x$ and $y$ indices refer to crystal axis $[100]$ and $[010]$ and not to the nanostructure elongations axes $X=[1\underline{1}0$ and $Y=[110]$.
Since $[100]$ and $[010$ are equivalent (whereas $X=[1\underline{1}0$ and $Y=[110]$ are not) hence $p_x$ and $p_y$ contributions must be identical.

The point of presenting Fig.~\ref{orbitalsandhhlh} is twofold.
First, in simplified models, and contrary to atomistic approaches, one very often neglects contributions to the ground hole state other than from $p_x$ and $p_y$ orbitals.
Such approach is apparently doubtful here since the contribution from $d$-orbitals reaches a significant $1/4$ fraction of the entire hole-function charge density.
Second, analysis of only selected components can be somewhat misleading.
This is emphasized on Fig.~\ref{orbitalsandhhlh} (right) where $p$ atomic orbitals where combined to form ''atomic'' heavy-hole and light-hole combinations
(e.g. $\left|\frac{3}{2}\right>=\frac{1}{\sqrt{2}}\left(\left|p_x^\uparrow\right>+i\left|p_y^\uparrow\right>\right)$)
and the sum of their squared moduli over all atoms in the system is plotted as a function of shape-elongation $t$.
$\left|\pm\frac{3}{2}\right>$ clearly various with the elongation, whereas separated square moduli of $p_x$ and $p_y$ do not.
This reveals an apparent change of phases between $p_x$ and $p_y$ atomic orbitals there was simply not visible in straightforward $\left|p_{x,y}\right|^2$ plots.
Next, the $p_z$ contribution grows with the deformation, yet it is present ($1\%$) even for a cylindrical case contrary to assumptions made by simplified approaches.
Moreover, the $p_z$ orbital or the light-hole ''atomic'' component (constructed from atomic $p_x$,$p_y$ and $p_z$ orbitals shown on Fig.~\ref{orbitalsandhhlh} (right))
reveals only a weak trend with the elongation. Thus the analysis of $p_z$ orbitals or light-hole ''atomic'' components only, and the neglect of the overall wave-function character (i.e. spatial charge density distribution) is not sufficient by itself to explain the increase of dark-exciton optical activity by nearly four orders of magnitude.

\bibliography{dashes}
\end{document}